\documentclass[12pt,twoside]{article}
\usepackage{fleqn,espcrc1}

\usepackage{graphicx}
\usepackage[figuresright]{rotating}

\newcommand{\AmS}{{\protect\the\textfont2
  A\kern-.1667em\lower.5ex\hbox{M}\kern-.125emS}}

\title{Antikaons in nuclei and dense nuclear matter}

\author{A.~Ramos\address{
Departament d'Estructura i Constituents de la Mat\`eria,
Universitat de Barcelona, Diagonal 647, 08028 Barcelona, Spain
},
S.~Hirenzaki\address{Department of Physics, Nara Women's University,
Nara 630-8506, Japan
},
S.S.~Kamalov\address{Laboratory of Theoretical Physics, JINR Dubna,
141980 Moscow region, Russia},
T.T.S. Kuo\address{Department of Physics and Astronomy, State University
of New York, Stony Brook, NY 11794, USA
},
Y.~Okumura$^{\rm b}$,
E.~Oset\address{
Departamento de F\'{\i}sica Te\'orica and IFIC,
Institutos de Investigaci\'on de Paterna, Aptdo. Correos 22085,
46071 Valencia, Spain},
A.~Polls$^{\rm a}$,
H.~Toki\address{Research Center for Nuclear Physics (RCNP), Osaka
University, Ibaraki, Osaka 567-0047, Japan
}
and
L.~Tol\'os$^{\rm a}$
}
       
\begin{document}

\maketitle

\begin{abstract}
We present recent progress on the properties of antikaons in nuclei and
dense nuclear matter as obtained from two ${\bar K}N$ interaction models:
one based on the lowest-order meson-baryon chiral lagrangian and the
other derived from a meson-exchange picture.
\end{abstract}

\section{Introduction}

Understanding the properties of kaons and antikaons 
in the nuclear medium is of especial relevance in the  
analysis of data from heavy-ion collisions, as well as in
the physics of neutron stars interiors. In the latter case, a
condensed phase of $K^-$ mesons might appear if they develop enough
attraction in the medium, such that it
becomes more favorable to neutralize the positive proton charge with antikaons
rather than with electrons
\cite{KN86,Brown92,Fujii96,Li97,Knorren95a,GS98L,Pons00}. Analysis of
kaonic atom data, sensitive to low densities, favor a strongly attractive
$K^-$ nucleus interaction~\cite{FGB94,gal00},
although recently weaker potentials have also been shown to describe the
data appropriately \cite{oku99,baca00}. 
Information on the higher density regime can be inferred
from the analysis of heavy-ion collisions data. For instance, 
the enhancement
of the $K^-$ yield measured recently at
GSI \cite{kaos97} can be explained by a strong
attraction in the medium for the K$^-$
\cite{cassing97,LLB97}, although
an alternative mechanism, based on the
in-medium enhanced $\pi\Sigma\to K^- p$
reaction, has also been suggested \cite{SKE99}.

Although all experimental indications point at an
attractive antikaon potential in the nuclear medium, the size
of this attraction has not really been well constrained by the data. It is
therefore interesting to explore what the theoretical models predict.
A traditional perspective is that of mean field approaches,
built within the modern framework of chiral lagrangians
\cite{LLB97,CHL96b,Mao99} or based on the relativistic Walecka
model \cite{Sch97}. However,
an especially useful point of view is that provided by microscopic
models \cite{alberg76,Koch94,WKW96,Lutz98,RO99,laura},
since they establish a bridge between the properties of the
${\bar K}N$ interaction in the medium and those in free space. This is
the approach followed in the present contribution, where we will focus on
the results obtained with two different
${\bar K N}$ interactions: one is taken from the modern chiral
meson-baryon lagrangian in $S$-wave \cite{AO}, and the other is
the meson-exchange model of the J\"ulich group \cite{julich}.

\section{${\bar K}N$ interaction in free space}

The dynamics of ${\bar K}N$ scattering is dominated by the presence of an
isospin zero $S$-wave resonance, the $\Lambda(1405)$. In potential models 
the ${\bar K}N$ scattering observables (${\bar K}=K^-$ or $\bar{K}^0$)
are derived from the scattering amplitude, obtained from the
Bethe-Salpeter equation
\begin{equation}
T_{ij} = V_{ij} + V_{il} G_{l} T_{lj} \ ,
\label{eq:BS}
\end{equation}
where the diagonal loop operator $G_l$ stands for the intermediate meson-baryon
($MB$) propagator and $V$ is a suitable $MB \to M^\prime B^\prime$
transition potential.

A connection with the chiral lagrangian was established in
ref.~\cite{NK}, where the
properties of the strangeness $S = -1$ meson-baryon sector were studied in
a
potential model that, in Born approximation,
had the same $S$-wave scattering length as the chiral lagrangian,
including both the lowest order and the momentum dependent $p^2$ terms.
Fitting five parameters, the scattering observables were reproduced and 
the $\Lambda (1405)$ resonance was generated dynamically from a 
coupled-channels Lippmann-Schwinger equation which included the
meson-baryon
states opened at the ${\bar K}N$ threshold.
In ref.~\cite{AO} it was shown that the data could be reproduced
using only the lowest order lagrangian and one
parameter, the cut-off used  to regularize the loop $G_l$.
The value of the cut-off, $q_{\rm max}=630$ MeV, was
chosen to reproduce the $K^- p$ scattering branching ratios
at threshold.
The essential
difference with respect to ref.~\cite{NK} was the inclusion  in
ref.~\cite{AO} of all ten meson-baryon
states ($K^- p$, $\bar{K}^0 n$,  $\pi^0 \Lambda$, $\pi^+ \Sigma^-$, $\pi^0
\Sigma^0$, $\pi^-\Sigma^+$, $\eta\Lambda$, $\eta\Sigma^0$, $K^+\Xi^-$ and $K^0
\Xi^0$) that can be generated from the octet of
pseudoscalar mesons and the octet of ground-state baryons,
the $\eta\Lambda$ state being the most relevant of the new included ones.
The success of this method is analogous to that
obtained in the meson-meson sector \cite{oller97}, an explanation of which
is found by applying the Inverse Amplitude Method in coupled
channels taking the lowest and next-to-lowest order meson-meson
lagrangians \cite{oller99}. 
A more recent work corroborates the success of
the lowest order meson-baryon lagrangian in reproducing the ${\bar K}N$
scattering observables from the point of subtracted dispersion relations
\cite{oller00}.


We will also explore the antikaon properties as obtained from the ${\bar K}N$
J\"ulich interaction \cite{julich}, which is based on a meson-exchange picture.
Most of the parameters were taken from their $KN$
potential in the strangeness $S=1$ sector and the few free ones left
were adjusted to reproduce the 
$K^- p$ elastic and inelastic cross section data at low energies. 
The advantage
of this interaction is that it contains all partial waves and allows for an
investigation of their effect on the antikaon potential at the high momenta
explored by heavy-ion collisions.

Although both ${\bar K}N$ interactions describe 
reasonably well the low energy $K^- p$ elastic and inelastic cross
sections, some of the
scattering observables at threshold, summarized in Table
\ref{table:1}, are not well reproduced. Clearly, some improvements can
still be made for the J\"ulich ${\bar K}N$
interaction since the branching ratios were not used for
adjusting the parameters of the model. 

\begin{table}[htb]
\caption{\small
${\bar K}N$ scattering observables at threshold}
\label{table:1}
\newcommand{\m}{\hphantom{$-$}}
\newcommand{\cc}[1]{\multicolumn{1}{c}{#1}}
\renewcommand{\tabcolsep}{1pc} 
\renewcommand{\arraystretch}{1.2} 
\begin{tabular}{@{}lccc}
\hline
         & Chiral \cite{AO} & J\"ulich \cite{julich} & EXP 
\cite{To71,No78,Adm81} \\
\hline
$\frac{\Gamma (K^- p \rightarrow \pi^+ \Sigma^-)}
{\Gamma (K^- p \rightarrow \pi^- \Sigma^+)}$ & 2.32 & 5.78  & $2.36 \pm 0.04$
\\
$\frac{\Gamma (K^- p \rightarrow \hbox{\small charged
particles)}}
{\Gamma (K^- p \rightarrow \hbox{\small all)}}$ & 0.627 & 0.600 &  $0.664 \pm 0.011$
\\
$\frac{\Gamma (K^- p \rightarrow \pi^0 \Lambda)}
{\Gamma (K^- p \rightarrow \hbox{\small neutral states})}$ & 0.213 & 0.447 &$0.189
\pm 0.015$\\
$a^{I=0}$ (fm) & $-1.93 + i\,1.68$ & $-1.71 + i\,1.28$ & $-1.70 + i\,0.68$ \\
$a^{I=1}$ (fm) & \m$0.52 + i\,0.51$ & \m$1.07 + i\,0.71$ & \m$0.37 + i\,0.60$
\\
\hline
\end{tabular}
\end{table}

\section{$K^-$ deuteron scattering}

The amplitudes of the chiral model were used in ref.~\cite{sabit00} to
study
$K^-$ scattering off the deuteron, the lightest nucleus. It is
well known that the impulse approximation fails in describing the $K^-$
scattering length and that more sophisticated approaches based on the solution
of the Fadeev equations are required \cite{chand,gal,torres}. 
In ref.~\cite{sabit00} the coupled-channels 
Fadeev equations 
were solved in the so-called fixed center approximation (FCA)
including the $K^- p n$ and the charge exchange $\bar{K}^0
nn$ channels, since those involving pions were found to give a small
contribution. The resulting $K^-$ deuteron scattering length is 
\begin{equation}
 A_{Kd}\,=\frac{M_d}{m_K+M_d}\int\,d{\bf r}\,\mid \varphi_d({\bf r})\mid^2\,
 {\hat A}_{Kd}(r)\ ,
\label{sk17}
\end{equation}      
with                     
\begin{equation}
 {\hat A}_{Kd}(r)=\frac{{\tilde a}_p + {\tilde a}_n +(2{\tilde
 a}_p{\tilde a}_n-b_x^2)/r - 2b_x^2{\tilde a}_n/r^2}{1-{\tilde a}_p
 {\tilde a}_n/r^2 + b_x^2{\tilde a}_n/r^3}\ ,
\label{sk19}
\end{equation}                                           
where ${\tilde a} = a\,(1+m_K/m_N)$, and 
$b_x={\tilde a}_x/\sqrt{1+{\tilde a}_n^0/r}$ is the charge
exchange amplitude renormalized due to the ${\bar K}^0 n$
rescattering. 
Expanding Eq.~(\ref{sk19}) to different orders in $(1/r)$
gives rise to the different approximations, such as the impulse
approximation (IA) (order zero) or the IA plus double
rescattering (first order).
In Table \ref{table:2} we collect the results of ref.~\cite{sabit00} 
obtained from the chiral elementary ${\bar K}N$ amplitudes
calculated using the physical basis or the isospin basis at an energy
$W=m_K+m_N = 1431.6$ MeV.

\begin{table}[htbp] \caption {\small
$K^-$-deuteron scattering length (in fm)
calculated using different approximations }
\label{table:2}
\newcommand{\m}{\hphantom{$-$}}
\newcommand{\cc}[1]{\multicolumn{1}{c}{#1}}
\renewcommand{\tabcolsep}{1pc} 
\renewcommand{\arraystretch}{1.2} 
\begin{tabular}{@{}ccc}
\hline
& Physical basis    & Isospin basis \\
\hline IA
& $ -0.260  + i\, 1.872 $ & $-0.318 + i\, 2.013 $ 
\\
\hline IA + double resc.
& $ -2.735  + i\, 2.895 $ & $-3.168 + i\, 3.717 $ 
\\
\hline $A_{Kd}$ (only el.resc.)
& $  -1.161 + i\, 1.336 $ & $-1.255 + i\, 1.518 $ 
\\
$A_{Kd}$ (total)
& $  -1.615 + i\, 1.909 $ & $-1.909 + i\, 2.455 $ 
\\                                                                               
\hline
\end{tabular}
\end{table}

We first observe that the results from the lowest orders differ
appreciably from those obtained by summing the full Fadeev series
in the FCA given by Eqs.~(\ref{sk17}), (\ref{sk19}). Secondly, 
the $K^-$ deuteron scattering length obtained from scattering amplitudes that
rely on isospin symmetry differ appreciably from that obtained within
the physical basis.
Finally, the result $A_{Kd}=-1.62 +i\, 1.91$ fm differs from
previous multichannel Fadeev approaches, $A_{Kd}=-1.47+i\, 1.08$ fm \cite{gal}
and $A_{Kd}=-1.34 +i\, 1.04$ fm \cite{torres}, especially for the
imaginary part. As observed in ref.~\cite{sabit00},
this is mainly due to the different
elementary amplitudes used in the different works. 
In fact, taking the amplitudes of ref.~\cite{torres} and using isospin
symmetry, which leads to a particularly small
imaginary part for the charge
exchange $K^-p \to \bar{K}^0 n$ amplitude,
one obtains the FCA result  $A_{Kd}=-1.54 +i\, 1.29$ fm, very close to
that
quoted in ref.~\cite{torres}. The studies in ref.~\cite{deloff}, together
with the
analysis performed in ref.~\cite{sabit00},  show that one can associate
the remaining differences to the FCA which
overestimates the full Fadeev calculation by about 15\%.

Although the results described here show that it is not simple to extract
the elementary scattering amplitudes from the deuteron data, it
can nevertheless be expected  
that the experimental results from the DEAR
experiment at Frascati will introduce a further check of consistency between
elementary amplitudes and should bring some light into issues like
chiral symmetry and partial isospin breakup.

\section{${\bar K}$ in nuclear matter}

One source of medium modification of the ${\bar K}N$ amplitude is that
induced by Pauli-blocking \cite{Koch94,WKW96} which effectively shifts
the intermediate
${\bar K}N$ states, and in turn the $\Lambda(1405)$ resonance, to higher
energies. Due to the strong energy dependence, this shift changes the
${\bar K}N$ amplitude around
threshold from being repulsive in free space to being attractive in the medium.
The ${\bar K}$ meson thus develops an attractive potential which, when
incorporated in
the equation determining the in-medium ${\bar K}N$ amplitude, compensates the
repulsive effect of Pauli blocking. When these effects are considered
self-consistently, it is observed that the position of the $\Lambda(1405)$ remains
unchanged \cite{Lutz98}. Since the $\pi \Sigma$ states have a strong influence in
the dynamics of ${\bar K}N$ scattering, the dressing of pions was also considered in
the study of ref.~\cite{RO99}, together with mean-field potentials for the baryons
participating in the coupled-channels problem.
These medium effects are incorporated by replacing the free baryon and
meson propagators by dressed ones in the meson-baryon loop-function 
which then reads
\begin{eqnarray}
G(P^0,\vec{P},\rho) &=&
\int_{\mid\vec{q}\,\mid < q_{\rm max} } \frac{d^3 q}{(2 \pi)^3}
\frac{M}{E (-\vec{q}\,)}
\int_0^\infty d\omega \,
 S(\omega,{\vec q},\rho) \nonumber \\
&\times & \left\{
\frac{1-n(\vec{q}_{\rm lab})}{\sqrt{s}- \omega
- E (-\vec{q}\,)
+ i \epsilon} +
\frac{n(\vec{q}_{\rm lab})}
{\sqrt{s} + \omega - E(-\vec{q}\,) - i \epsilon } \right\} \ ,
\label{eq:gmed}
\end{eqnarray}
where $(P^0,\vec{P})$ is the total four-momentum in the lab frame,
$n(\vec{q}_{\rm lab})$ is the nucleon occupation probability, and
$S(\omega,\vec{q},\rho)=-{\rm Im}
D(\omega,\vec{q},\rho)/\pi$ is the ${\bar K}$ spectral density,
which in free space reduces to
$\delta(\omega-\omega(\vec{q}\,))/2\omega(\vec{q}\,)$.
The in-medium antikaon propagator,
$D(\omega,\vec{q},\rho)$,
is determined from the antikaon self-energy, $\Pi_K(\omega,\vec{q},\rho)$,
which is obtained by summing the
in-medium $\bar{K}N$ interaction, $T_{\rm
eff}(P^0,\vec{P},\rho)$,
over the nucleons in the Fermi sea
\begin{equation}
\Pi_K(q^0,{\vec q},\rho)=2\sum_{N=n,p}\int
\frac{d^3p}{(2\pi)^3}
n(\vec{p}\,) \,   T_{\rm eff}(q^0+E(\vec{p}\,),\vec{q}+\vec{p},\rho) \ .
\label{eq:selfka}
\end{equation}
Note that a self-consistent approach is required since one
calculates the ${\bar K}$ self-energy from the effective
interaction $T_{\rm eff}$ which uses ${\bar K}$ propagators which
themselves include the self-energy being calculated.

\begin{figure}[htb] 
\begin{center}
\includegraphics[width=0.6\linewidth]{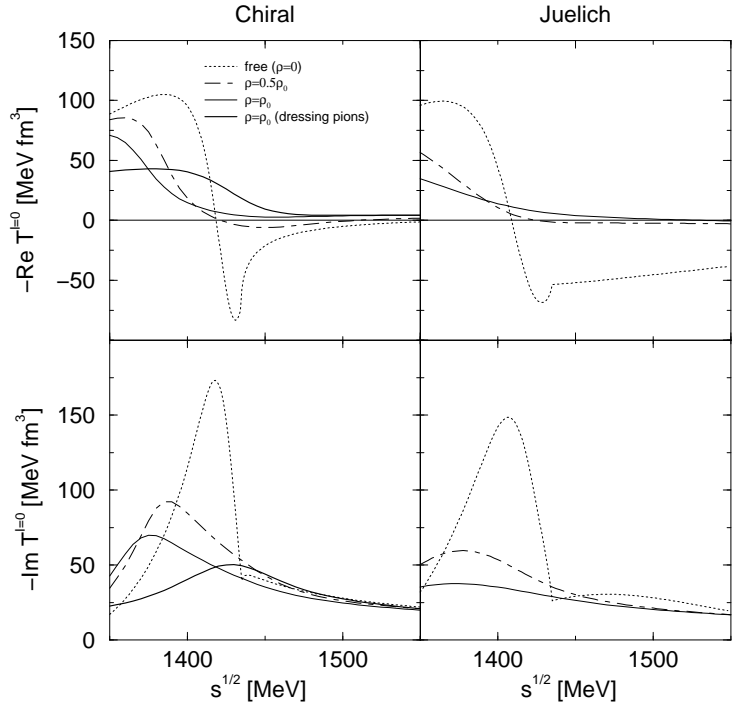}
\end{center}
\caption{\small
Real and imaginary parts of the $I=0$ ${\bar K}N$
scattering amplitude as
functions of $\sqrt{s}$ for $\mid \vec{p}_K + \vec{p}_N\mid = 0$ and
several densities (from \cite{RO99} and \cite{laura}).}
\label{fig:ampl}
\end{figure}

The free ${\bar K}N$ amplitude in the $I=0$, $L=0$ channel
is shown in Fig.~\ref{fig:ampl}, where it is compared  
with the in-medium one
at two nuclear densities, $\rho=\rho_0=0.17$ fm$^{-1}$
and $\rho=0.5\rho_0$. Results are shown for two different
calculations, that
of ref.~\cite{RO99},
based on the chiral lagrangian model \cite{AO}, and that of
ref.~\cite{laura}, based on the
meson-exchange ${\bar K}N$ interaction \cite{julich}.
We observe that the medium modified amplitudes show the same
qualitative trends. Note also how the real part of the amplitude
(upper panels) at the $K^- p$ threshold
($\sqrt{s}=1433$ MeV) is repulsive in free space and attractive in the
medium. The thick solid line on the left panels show the additional effect
on the ${\bar K}N$ amplitude of
dressing the pions in the intermediate states \cite{RO99}.

Most of the available models study the in-medium ${\bar K}N$
amplitude in $S$-wave.
However, if one aims at extracting the properties of antikaons
through the
analysis of heavy-ion collisions, one must keep in mind that they are
created at a finite momentum of around $250-500$ MeV/c, hence the effect
of higher partial waves might be relevant. The meson-exchange
${\bar K}N$ potential of the
J\"ulich group \cite{julich} is given in partial waves and 
allows a straightforward analysis of the importance of the $L>0$
components. From the chiral perspective, the $P$-wave amplitudes up to the
next-to-leading order ${\bar K}N$ chiral lagrangian have been identified
and the parameters have been fitted to reproduce a large amount
of low energy data \cite{caro}. However, a nuclear medium application of this model
is not available yet.

\begin{figure}[htb]
\begin{minipage}{.38\linewidth}
\begin{center}
\includegraphics[width=0.85\linewidth]{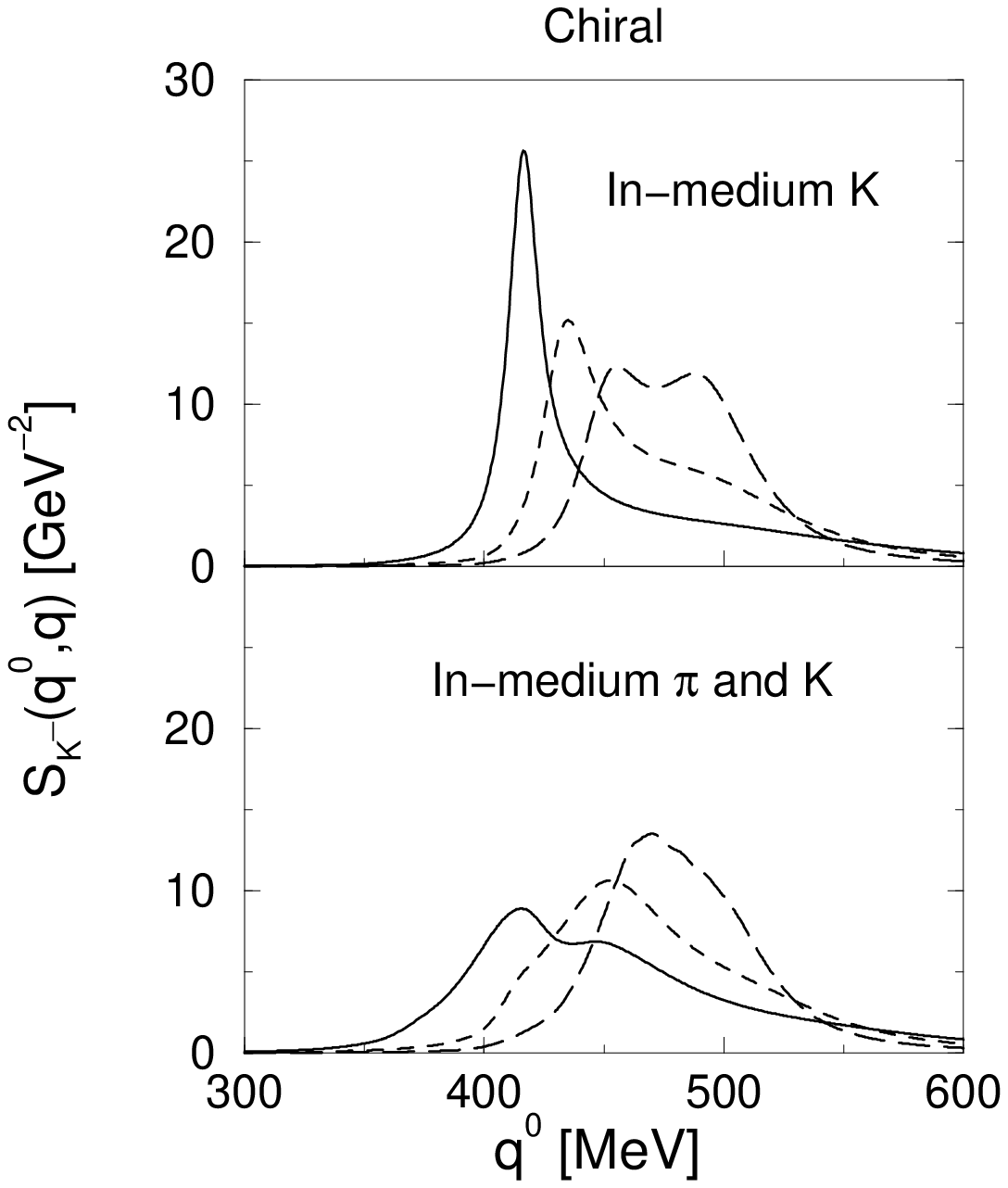}
\end{center}
\caption{\small
$K^-$ spectral density for zero momentum from the
chiral model of ref. \cite{RO99}}.
\label{fig:kspec}
\end{minipage}
\hfill
\begin{minipage}{.55\linewidth}
\begin{center}
\includegraphics[width=0.7\linewidth,angle=-90]{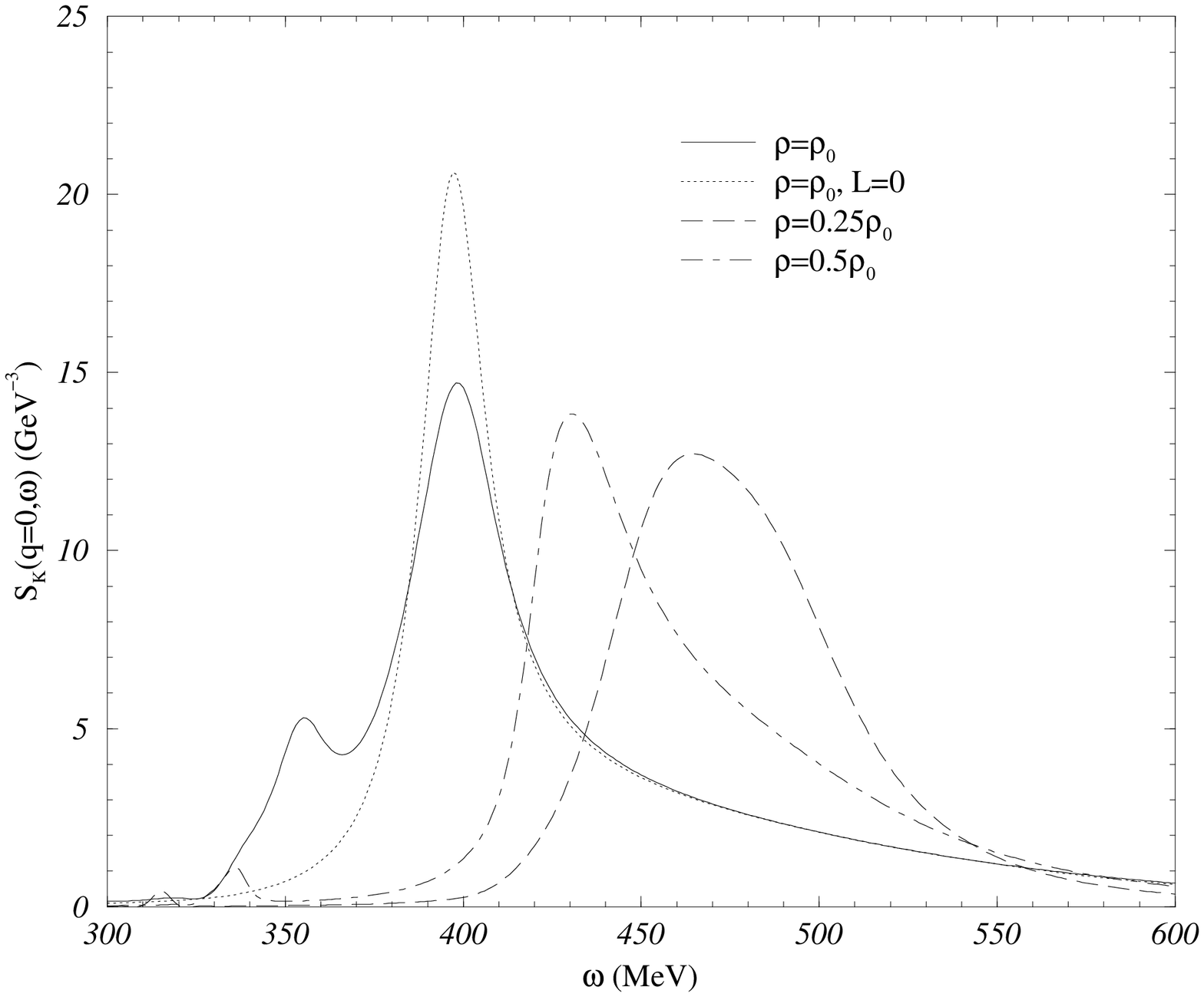}
\end{center}
\caption{\small
$K^-$ spectral density for zero momentum
  using the J\"ulich ${\bar K}N$ potential (from \cite{laura}).
 }
\label{fig:kspec2}
\end{minipage}
\end{figure}

As an example on the importance of dressing the antikaon, we show in 
Fig.~\ref{fig:kspec} 
the spectral function of a $K^-$ meson of zero
momentum obtained with the chiral model of ref.~\cite{RO99} for various
densities: $\rho_0$ (solid line), $\rho_0/2$ (short-dashed line) 
and $\rho_0/4$ (long-dashed line). The upper pannel corresponds to a
calculation in which the antikaons are dressed selfconsistently but the
pions are free. At $\rho_0/4$ one sees two excitation modes. The left one
corresponds to the
$K^-$ pole branch, appearing
at an energy smaller than the kaon mass, $m_K$, due to the
attractive ${\bar K}N$ amplitude. The peak on the right corresponds to
the $\Lambda(1405)$-hole excitation mode,
which appears around $m_K$.
As density increases, the $K^-$ feels an enhanced
attraction while
the $\Lambda(1405)$-hole peak loses
strength and dissolves in the dense nuclear medium. 
When the dressing of the pion is also incorporated, the effective
interaction
$T_{\rm eff}$ becomes even smoother and
the resulting
$K^-$ spectral function is displayed in the bottom panel of
Fig.~\ref{fig:kspec}. Even at
very small densities one can no longer distinguish the
$\Lambda(1405)$-hole
peak from the $K^-$ pole one.
As density increases, the attraction
felt by the $K^-$ is more moderate and the $K^-$ pole peak
appears at
higher energies than before.
However, more strength is found at very
low energies, especially at $\rho_0$, due to the particle-hole
($ph$) components
of the pionic strength, which couple the
${\bar K}N$ pair to 
$ph\Sigma$ states.
It is precisely the opening of the
$\pi\Sigma$ channel, on top of the already opened $ph\Sigma$ one,
which causes a cusp around 400 MeV.
The calculation of ref.~\cite{laura} using the
J\"ulich ${\bar K}N$ interaction, shown in
Fig.~\ref{fig:kspec2},
 obtains qualitatively similar
results as those in the upper panel of Fig.~\ref{fig:kspec}, where only
antikaons are dressed.
We notice an additional structure in the spectral
function to the left of the quasiparticle peak at energies
of the ${\bar K}$ around $320-360$ MeV, which is not present when only
the $L=0$ component of the ${\bar K}N$ interaction is retained
(dotted line). This peak is indicating the physical in-medium excitation
of $\Sigma h$ states with antikaon quantum numbers coming from the
$L=1$, $I=1$ component of the ${\bar K}N$ interaction.

A non-relativistic  antikaon potential can be obtained from
the self-energy
at the quasiparticle energy via the relation
\begin{equation}
U_K(\vec{q}\,)=
\frac{\Pi_{K}(\varepsilon_{qp}(\vec{q}\,),{\vec q},\rho)}{2 m_K} \ .
\label{eq:upot}
\end{equation}

\begin{figure}[h!]
\vspace*{-1cm}
\begin{minipage}{.45\linewidth}
\centerline{
     \includegraphics[width=0.75\textwidth]{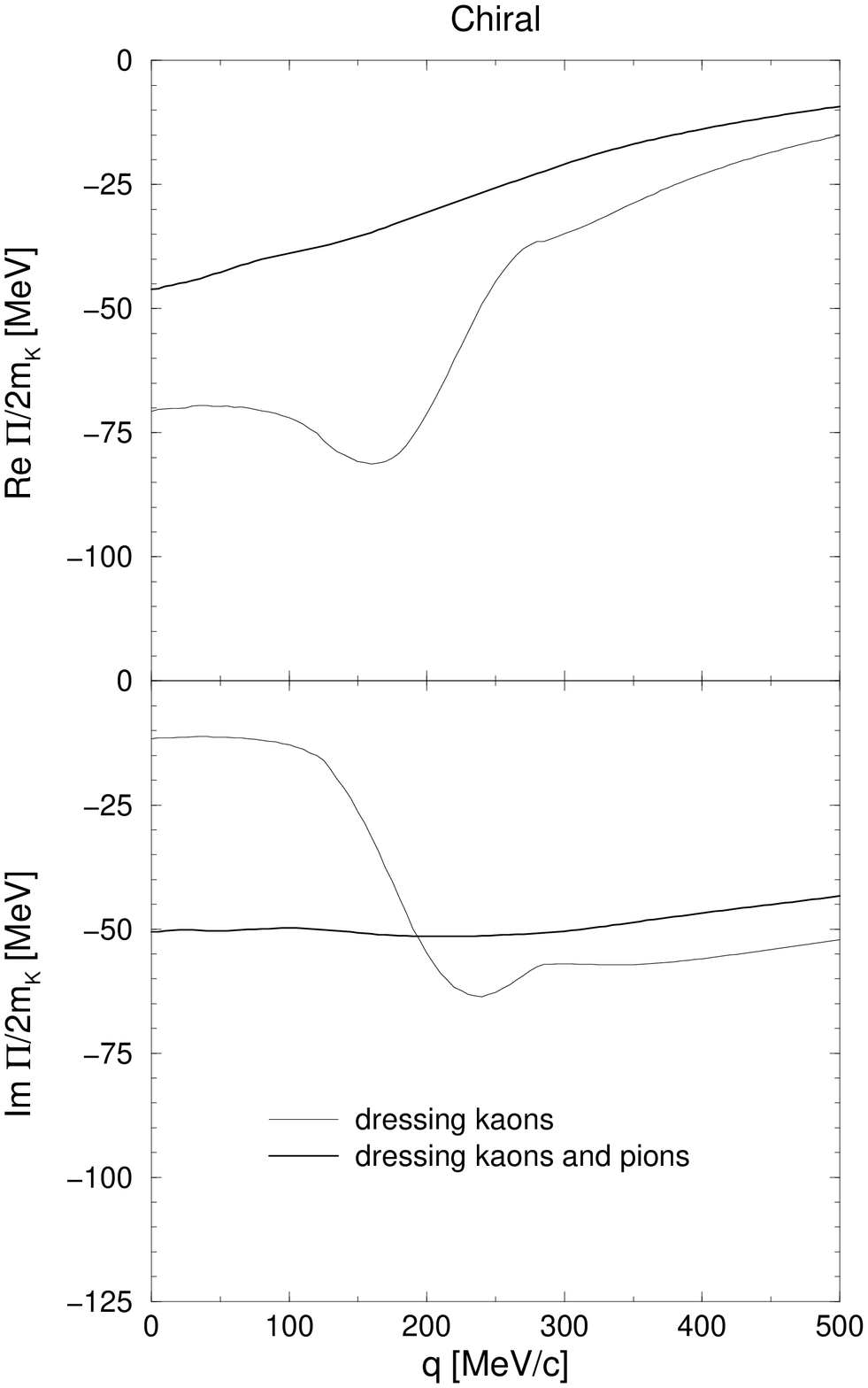}
}
      \caption{\small
Real and imaginary parts of the ${\bar K}$ optical potential
at $\rho=\rho_0$ as functions of the antikaon momentum obtained from
the chiral model of ref.~\cite{RO99}.}
        \label{fig:upot1}
\end{minipage}
\hfill
\begin{minipage}{.45\linewidth}
\centerline{
     \includegraphics[width=0.75\textwidth]{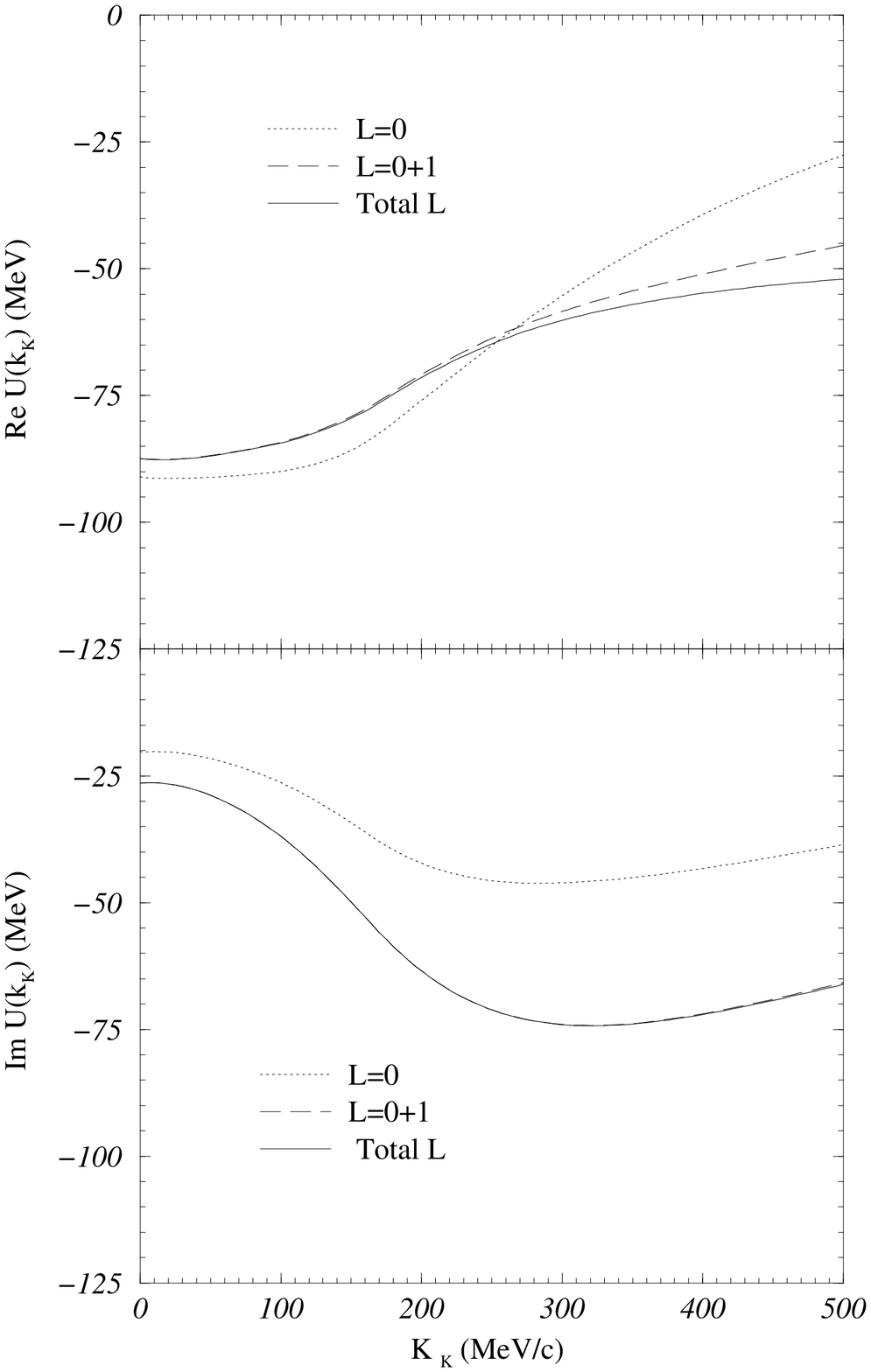}
}
      \caption{\small
The same as Fig.~\ref{fig:upot1}
obtained in ref.~\cite{laura} using the J\"ulich
 ${\bar K}N$ potential.}
        \label{fig:upot2}
\end{minipage}
\end{figure}

The momentum dependence of the potential obtained
from the chiral model of ref.~\cite{RO99} at $\rho=\rho_0$ is shown in
Fig.~\ref{fig:upot1} for two approximations, one in which only
the antikaons are dressed self-consistently
(thin solid lines) and another in which the self-energy of the
pions is also included (thick solide line).
Note that the antikaon potential obtained when
the pions are also dressed has much less structure. This is due in part 
to the
smoother in-medium amplitude, but also to the different quasiparticle
energy at which the antikaon self-energy is evaluated. This quasiparticle
energy is more attractive when only antikaons are dressed and, hence,
the amplitude is explored at lower energy regions, closer to the position
of the in-medium $\Lambda(1405)$ resonance.
Results from the model of ref.~\cite{laura} using the J\"ulich ${\bar
K}N$ potential are shown in Fig.~\ref{fig:upot2}, where
the effect of including the higher partial waves can be seen.
Already at zero momentum, Fermi motion induces some small
contribution of partial waves
higher than $L=0$. Clearly, the effect of the
higher partial waves increases with increasing ${\bar K}$ momentum,
flattening out the real part of the potential and producing more
structure to the imaginary part. At an antikaon momentum of around 500
MeV/c, the inclusion of the higher partial waves practically doubles the
size of the antikaon potential with respect to the $S$-wave value.

\section{Kaonic atoms}

\begin{figure}[ht!] 
\begin{center}
\vspace*{-1cm}
\includegraphics[width=0.45\linewidth]{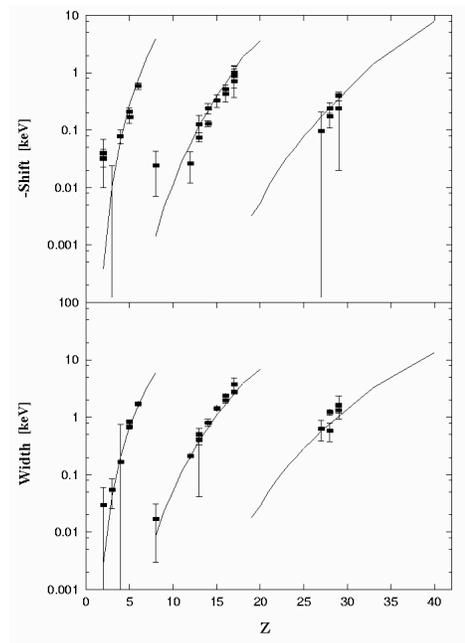}
\end{center}
\vspace*{-1cm}
\caption{\small
Energy shifts and widths of kaonic atom states
(from
 \cite{oku99}). The experimental data are taken from
the compilation given in \cite{BGF97}.}
\label{fig:katom}
\end{figure}

Since in kaonic atoms the $K^-$ is bound with a small
(atomic) energy, their
study requires the knowledge of the
antikaon self-energy at
$(q^0,\vec{q}\,)=(m_K,\vec{0}\,)$.
The density-dependent self-energy obtained with the chiral model
has been recently used to study kaonic atoms
in the framework of a local density approximation \cite{oku99}, which
amounts to
replace the nuclear matter density $\rho$
by the density profile $\rho(r)$ of the particular nucleus.
The results, displayed in Fig.~\ref{fig:katom}, show that
both the energy shifts
and widths of kaonic atom states agree well with the
bulk of experimental data \cite{BGF97}. Recent investigations that
respect the low density theorem \cite{carmen00} show that the
non-local corrections, coming from the energy and
momentum dependence of the potential,
from the $P$-wave terms of the meson-baryon
lagrangian and from Fermi motion, are small 
and lead to changes in the widths and shifts
smaller than the experimental errors.

Reproducing kaonic atom data with the moderately attractive
antikaon nucleus potential of $-45$ MeV obtained from the chiral model
\cite{oku99} is
in contrast with
the depth of around $-200$ MeV obtained
from a best fit to $K^-$ atomic data with a phenomenological
potential that includes an additional
non-linear density
dependent term \cite{FGB94}.
A clarifying quantitative comparison of kaonic atom results obtained
with various $K^-$-nucleus potentials can be found in ref.~\cite{baca00}.
It is shown there that adding to the chiral potential a
phenomenological piece, which is fitted to the data, the
$\chi^2/d.o.f.$ is reduced from 3.8 to 1.6. 
The resulting potential is slightly more attractive ($-50$ MeV at
$\rho_0$) and the imaginary part is reduced by about a factor 2. The work
of ref.~\cite{baca00} reemphasizes
that kaonic atoms only explore the antikaon potential at the
surface of the nucleus. Therefore,
although all models predict attraction for
the $K^-$-nucleus
potential, the precise
value of its depth at the center of the nucleus, which has important
implications for the occurrence of kaon condensation, is still not known.
It is then necessary
to gather more data that could help in disentangling the
properties of the $\bar{K}$ meson in the medium. Apart from the valuable
information that can
be extracted from the production of $K^-$
in heavy-ion collisions, one could also study deeply bound
kaonic states,
which have been predicted to be narrow \cite{oku99,baca00,FG99a} and
could be measured
in $(K^-,\gamma)$ \cite{oku99} or $(K^-,p)$ reactions
\cite{FG99b,Kishi99}.

\section*{Acknowledgments}
We would like to acknowledge financial support from the DGICYT (Spain) 
under contracts PB96-0753, PB98-1247, from the Generalitat de Catalunya
under grant SGR2000-24, and from the EU TMR network Eurodaphne contract
ERBFMRX-CT98-0169. L.T. acknowledges support from a doctoral fellowship of
the Ministerio de Educaci\'on y Cultura (Spain).

\end{document}